\documentclass{jjap2}
\usepackage{subfigure}
\usepackage{here}
\usepackage{multirow}
\title{Trace Element Analysis of Potassium by \\Resonance Ionization Mass Spectrometry}

\author{Yoshihiro \textsc{Iwata}$^{}$\thanks{E-mail: yiwata@icepp.s.u-tokyo.ac.jp}, Yoshizumi \textsc{Inoue}$^{1}$, and Makoto \textsc{Minowa}$^{}$}

\inst{$^{}$Department of Physics, School of Science, University of Tokyo, 7-3-1 Hongo, Bunkyo-ku, Tokyo 113-0033\\
$^{1}$International Center for Elementary Particle Physics (ICEPP), University of Tokyo, 7-3-1 Hongo, Bunkyo-ku, Tokyo 113-0033}

\abst{A simple resonance ionization mass spectrometer is built with a
Quadrupole Mass Spectrometer (QMS) and two laser diodes aiming at trace
analysis of potassium.
The overall detection efficiency is estimated comparing the yields of
resonance ionization and electron-impact ionization in the same QMS.
A clear discrimination of $^{39}\rm K$, $^{40}\rm K$ and $^{41}\rm K$ is 
demonstrated with a help of isotope shifts of the atomic levels.}

\kword{RIMS, trace element analysis, potassium, graphite crucible, atom
beam, isotope discrimination, contamination assessment, semiconductor
wafer}

\begin{document}
\maketitle

\section{Introduction} 
Recently, the evaluation technology of the trace substances has 
become important in many areas such as environmental science, biochemistry 
and semiconductor industry. As one of the effective methods for the 
trace element analysis, Resonance Ionization Mass Spectrometry (RIMS) 
is currently studied mainly with calcium (Ca) or strontium (Sr) atoms 
for isotope analysis.\cite{ca-ex,sr-ex}
RIMS consists of two parts which are photoionization of a particular 
element by a tuned laser, and isotope discrimination by a conventional 
mass spectrometer like a Quadrupole Mass Spectrometer (QMS) or a 
Time-Of-Flight Mass Spectrometer (TOF-MS).
The first part is resonance excitation of specific atoms with a 
monochromatic laser followed by ionization with the same or another laser.
It is free from isobaric interference often seen in other methods 
such as Inductively-Coupled Plasma Mass Spectrometry (ICP-MS).
Additional isotopic selectivity can also be achieved with a CW laser
diode because of its narrower linewidth than the slight isotope shifts
of the resonance excitation wavelength. 

In this paper, we report on a demonstration of the trace element 
analysis of potassium gas atoms performed by a simple resonance 
ionization mass spectrometer built with two laser diodes and a Quadrupole 
Mass Spectrometer (QMS). 
The laser diode is used for the resonance excitation because the 
wavelength is tunable, although the variable range is narrow, and less 
expensive than other kinds of wavelength-variable lasers like a dye laser.
The potassium atom beam was prepared by decomposition of $\rm K_2CO_3$ in 
an electrothermally-heated graphite crucible. 

Difference of the resonance excitation wavelength between $^{39}{\rm K}$
and $^{41}{\rm K}$ was observed together with the mass peak of the 
naturally-occurring radioactive isotope $^{40}{\rm K}$ by fine-tuning 
the laser wavelength for resonance excitation.

The advantage of RIMS over the conventional contamination assessment
on the surface of semiconductor wafers by ICP-MS is described 
in terms of insensitivity to isobaric interferences and lower noise 
level of the channeltron detector.

\section{Experimental Setup}
\subsection{RIMS scheme}
\label{scheme}
\begin{figure}[H]
\begin{center}
\includegraphics[height=7cm]{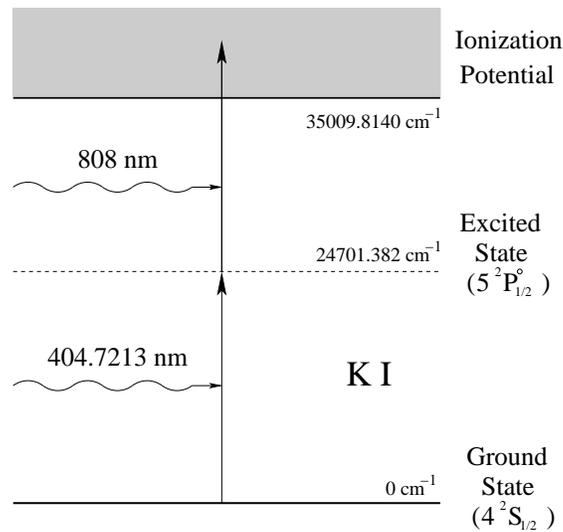}
\end{center}
\caption{Resonance ionization scheme of potassium used in this work.}
\label{ris-scheme}
\end{figure}

As mentioned above, RIMS method is the combination of the resonance
ionization by lasers and the mass analysis. Fig.~\ref{ris-scheme} shows the
resonance ionization scheme of potassium employed in the present study.
Two laser diodes were used for the single-resonance ionization of 
potassium gas atoms.

The first laser is a wavelength-tunable External Cavity Diode Laser 
(ECDL, Sacher Lasertechnik model SYS-100-405-20) operating in the 
wavelength range of $\lambda$=402.7$\--$405.2 nm. 
A single-mode $3.5\mu$m core diameter fiber is coupled to this laser 
system with an FC/APC connector.
The wavelength was fine-tuned to $\lambda=404.7213$ nm for the 
$4\,{^2{\rm S}_{1/2}}\,\rightarrow$\, $5\,{^2{\rm P}^\circ_{1/2}}$ 
resonance excitation in air,\cite{RIS-data} via the piezo actuator 
attached on the diffraction grating in the laser head. 
The piezo voltage can be controlled manually from the control panel 
knob in the range of $0\--100$ V, with a frequency sensitivity 
of $\Delta\nu=2$ GHz/V (0.0011 nm/V).
Higher piezo voltage means lower laser wavelength, or higher laser 
frequency.
The maximum laser output was more than $40$mW, corresponding to the 
single-mode fiber output of more than $20$mW. 
This laser has not been frequency-locked yet in this work, so the 
wavelength drift occurred in the time span of a few minutes or longer.

The second laser is an Amonics model ALD-808-3000-B-FC laser diode for 
photoionization with a fixed wavelength of $\lambda=808\pm5$nm. 
A multi-mode $100\mu$m core diameter fiber is coupled to this laser 
system with an FC/PC connector.
The maximum output was more than $3$W, corresponding to the multi-mode
fiber output of more than $2.5$W. 

We used a Pfeiffer Vacuum model QMS200 quadrupole mass spectrometer 
for the mass analysis of potassium ions produced by the resonance 
ionization. It is equipped with a channeltron to detect ions as an 
amplified ion current.

\subsection{Sample preparation}
In this work, potassium gas atoms were obtained from the 
decomposition of $\rm K_2CO_3$ in the electrothermally-heated 
graphite crucible. 
In each trial, $15\mu\ell$ of 50\% water solution of $\rm K_2CO_3$
was loaded into the crucible.
After drying the sample, the crucible containing the net weight of 
$14\pm1$ mg $\rm K_2CO_3$, corresponding to $1.2\times10^{20}$ 
potassium atoms, was set in the heating device 
(Epiquest model THKC-200-SB).

As the crucible temperature increases electrothermally, the 
decomposition is assumed to occur as follows:\cite{k2co3}
\begin{eqnarray}
{\rm K_2CO_3(c)} &\rightleftharpoons& {\rm K_2O(c)} + {\rm CO_2(g)}.
\label{decomp1}
\end{eqnarray}
The potassium atoms are obtained from the decomposition of ${\rm K_2O}$
under the higher crucible temperature: 
\begin{eqnarray}
{\rm K_2O(g)} &\rightleftharpoons& 2{\rm K(g)} + \frac{1}{2}{\rm O_2(g)}.
\end{eqnarray}
The number $N_{\rm K}$ of the obtained potassium gas atoms 
per unit time is
\begin{eqnarray}
N_{\rm K} = \frac{FP_{\rm K}}{k_BT},
\end{eqnarray}
where $F, P_{\rm K}, k_B$ and $T$ represent the crucible conductance, 
the partial pressure of K(g), Boltzmann constant, and the temperature 
inside the crucible, respectively. Assuming a molecular flow regime, 
\begin{eqnarray}
F = \frac{2\pi a^3\bar{v}}{3L},
\end{eqnarray}
where $a$ and $L$ are the crucible radius and length, and $\bar{v}$ is 
the average velocity of the potassium gas atoms. In the present 
experiments, $F=8.4\times10^{-5}$ [${\rm m^3}$/s] at $T=1174{\rm K}$. 
Combined with $P_{\rm K}=7.0\times10^{-6}$ [atm] at this temperature,
\cite{k2co3} $N_{\rm K}=3.6\times10^{15}$ [1/s] is expected. 
The potassium atoms were introduced into the QMS as an atom beam through 
an orifice of $\phi1$mm.

\subsection{Schematic view of the experimental setup}
\label{exp}
\begin{figure}[H]
\begin{center}
\includegraphics[height=7cm]{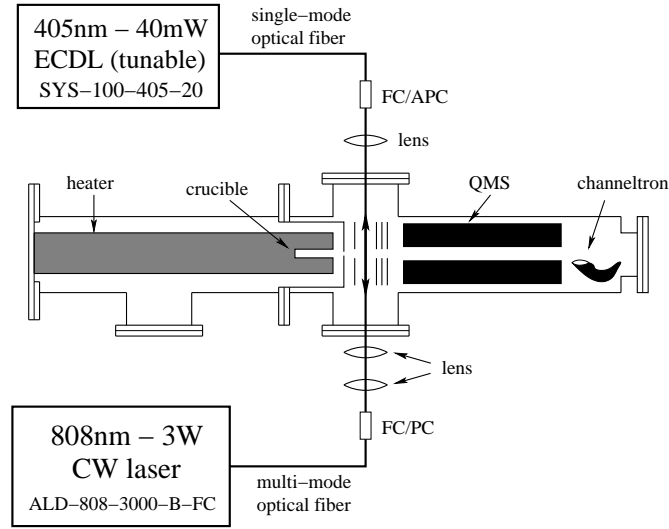}
\end{center}
\caption{Schematic view of the experimental apparatus. Two laser beams are 
irradiated perpendicularly to the direction of the potassium atom beam, and 
at an angle of 166.7$^\circ$ to each other.}
\label{setup}
\end{figure}

A schematic view of the experimental apparatus is shown in
Fig.~\ref{setup}. The potassium gas atoms from the crucible are 
resonantly-ionized mainly in the overlap region of two laser beams 
irradiated perpendicularly to the direction of the potassium atom beam.

The fiber output of each laser diode is focused to the ionization region 
in vacuum with Melles Griot model optical lenses.
The lens for the 405nm laser is a 01LUD009 Symmetric-Convex Precision 
Fused-Silica Singlet Lens with a focal length of 30.0mm.
Two lenses for the 808nm laser are a 06GLC006 Collimating lens 
with a focal length of 50.1mm and a 06LAI011 Diode Laser Glass
Doublets with a focal length of 100.0mm. 
These two lenses are corrected for spherical aberration in advance, 
so the focal spot of the 808nm laser is mainly determined geometrically 
to be a diameter of about $200\mu$m by the position of the lenses and 
the core diameter of the multi-mode fiber.
The 405nm laser spot is set to a diameter of about $500\mu$m so that 
the overlap region of two laser beams can be easily found.
To avoid any damage caused by the opposite laser beam coming through the
fiber, two laser beams are irradiated at an angle of 166.7$^\circ$ to 
each other.

The laser power at the ionization region is reduced to about $20$mW 
for 405nm laser and $1.5$W for 808nm laser, because of the 
transmission loss of the lenses and the viewport windows. 
Each optical fiber output is fixed to the viewport of the QMS 
to avoid misalignment of the laser spot due to vibration of the 
vacuum system.
Ions are extracted, mass analyzed by the QMS, and finally detected 
by the off-axis channeltron detector as shown in Fig.~\ref{setup}.

In the ionization region is placed a filament made of yttrium 
oxide coated iridium, which was switched on for Electron-Impact (EI)
ionization of the potassium atoms.
EI runs were performed as a reference for the estimatation of 
RIMS overall detection efficiency (see Section~\ref{ODE} for details).
It should be noted that any atoms or molecules can be ionized by EI, 
leading to the increase of the background noise of the channeltron 
current. The emission current of the thermal electron was set to 0.1mA 
in this work.

The crucible temperature was measured with a thermocouple on the 
outer bottom of the crucible. Based on the test data, the thermocouple
temperature indication was $40\--50$K higher than the real temperature 
$T$ inside the crucible in $T\sim1200$ K region.

\section{Results and Discussion}
\subsection{Resonance ionization signal}
\begin{figure}[H]
\begin{center}
\includegraphics[height=7cm]{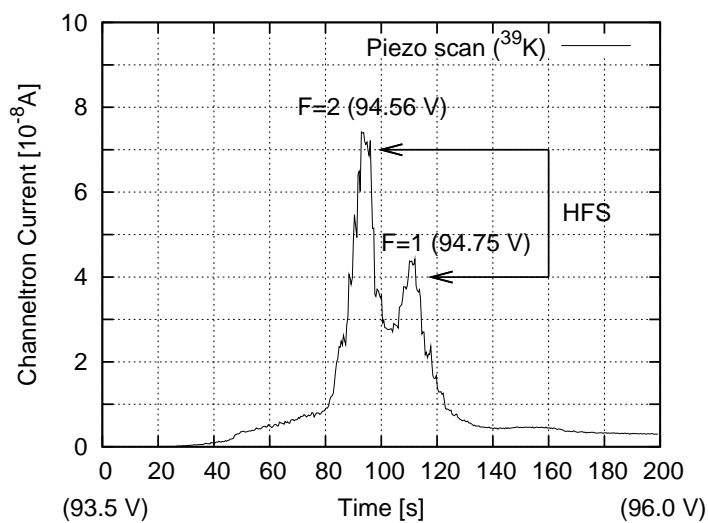}
\end{center}
\caption{An example of the resonance ionization signal of $^{39}\rm K$ by scanning the piezo voltage $V_{\rm piezo}$. The QMS ion current is shown as a function of time for the voltage scanning. The observed two peaks correspond to the hyperfine splitting (HFS) of $^{39}\rm K$\,\,$4\,{^2{\rm S}_{1/2}}$ ground state. This figure represents the same data as $^{39}\rm K$ data of Fig.~\ref{signal-all}.}
\label{signal}
\end{figure}

Resonance ionization signal is observed by fine-tuning the 
piezo voltage $V_{\rm piezo}$ to the resonance excitation wavelength.
Fig.~\ref{signal} shows an example of the observed resonance ionization 
signal of $^{39}\rm K$ by scanning the piezo voltage $V_{\rm piezo}$ 
manually from 93.5 V to 96.0 V. 
It should be noted that the laser wavelength changes with $V_{\rm piezo}$
as well as with the laser current or temperature, so only the difference 
of the piezo voltage $V_{\rm piezo}$ has meaning, but the value itself 
has no meaning.

The observed two peaks are consistent with the hyperfine splitting (HFS) 
of $^{39}\rm K$\,\,$4\,{^2{\rm S}_{1/2}}$ ground state. 
See Section~\ref{HFsignal} for the detailed explanation.

\subsection{Overall detection efficiency}
\label{ODE}
The overall detection efficiency is defined as the ratio of the 
number $N_{\rm D}$ of potassium atoms detected by the QMS to that 
initially loaded in the crucible.
$N_{\rm D}$ could be experimentally determined by the time integration of the 
channeltron current until the ${\rm K_2CO_3}$ sample in the crucible is 
evaporated away. 
Since we have not yet used the frequency-locking system for the laser
of the resonance excitation, it is difficult to keep for a long time 
the laser frequency on-resonance with the transition to the excited 
state shown previously in Fig.~\ref{ris-scheme}.
On the other hand, it is rather easy to estimate the overall detection 
efficiency $E_{\rm EI}$ of EI ionization experimentally by integrating 
the channeltron current of EI ionized potassium.
Therefore, RIMS overall detection efficiency $E_{\rm RIMS}$ in this work
can be determined by the short time measurement of the ratio $R$ of the 
detection efficiency by RIMS to EI: $E_{\rm RIMS}=R\times E_{\rm EI}$.
 
To estimate the detection efficiency $E_{\rm EI}$, the sum of the
channeltron current $I_{\rm Meas}$ of $^{39}\rm K^+$ and $^{41}\rm K^+$
has been measured from the beginning of the crucible heating ($t=92$ s) 
until the $\rm K_2CO_3$ sample was lost ($t=25900$ s).
It took about an hour to increase the crucible heating current from 
0 A ($t=92$ s) to 7.50 A, corresponding to the crucible inner temperature 
of $T\sim300$ K to $T\sim1170$ K.
After reaching the target termperature, the heating current was 
maintained at 7.50 A until the end of the measurement. 
Background channeltron current $I_{\rm BG}$ has been estimated as the 
average of the channeltron current without crucible heating 
($t=$ 0 -- 92 s). Integration of the net channeltron current 
$I_{\rm SG}\equiv I_{\rm Meas}-I_{\rm BG}$ over the time span of 
92 -- 25900 s yielded $5\times10^{-3}$ [C], correponding to 
$9\times10^{10}$ potassium atoms, which was 
$E_{\rm EI}\sim8\times10^{-10}$ of the initially loaded ones.

\begin{figure}[H]
\begin{center}
\begin{tabular}[b]{c}
\subfigure[Measurement of the channeltron current by EI for the
 estimation of the detection efficiency.]{\includegraphics[height=5cm]{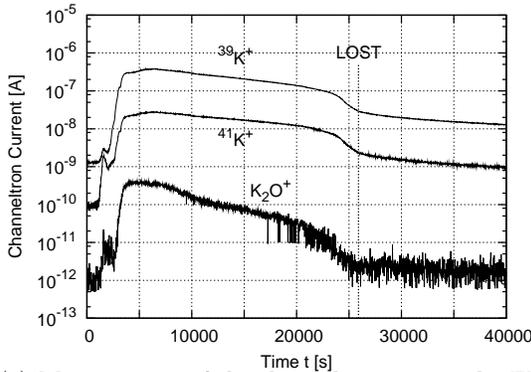}}
\hspace{0.8cm}
\subfigure[Comparison of the measured channeltron current between RIMS and
 EI. The channeltron voltage was set lower than (a). See text for the meaning of (I)--(IV).]{\includegraphics[height=5cm]{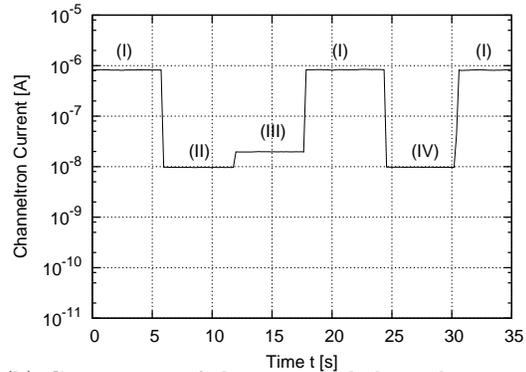}}
\end{tabular}
\end{center}
\caption{Measurement of the channeltron current by RIMS and
 Electron-Impact (EI) ionization for the estimation of the detection efficiency by each method. See text for details.}
\label{RIMS-DE}
\end{figure}

Fig.~\ref{RIMS-DE}-(b) shows an experimental result of the channeltron 
current of $^{39}\rm K^+$ atomic ions by RIMS and EI. 
The channeltron voltage in this figure was lower than that used in the 
measurement of Fig.~\ref{RIMS-DE}-(a). Therefore, the ion currents of 
Fig.~\ref{RIMS-DE}-(a) and Fig.~\ref{RIMS-DE}-(b) cannot be directly 
compared.

The crucible heating current was set to 7.50A in advance.
Throughout the measurement, the emission current of the thermal electron 
for EI ionization has been kept to 0.1mA, in other words, the
channeltron current contributed by EI ionization remained constant. 
In addition, two lasers were irradiated in 4 ways: (I) both were ON; 
(II) both were OFF; (III) 405nm laser was ON and 808 OFF; (IV) 405 OFF
and 808nm laser was ON.
Little difference was observed between the current of (II) and
(IV), meaning that resonance ionization occurs only under the existence
of the wavelength fine-tuned laser for resonance excitation.
The ratio $R$ of the detection efficiency by RIMS to EI can be obtained 
as the ratio of the measured channeltron current of (I) subtracted by (II) 
to (II), which was
$R\simeq8.3\times10^{-7}{\rm A}/9.7\times10^{-9}{\rm A}\sim90$.
Combined with $E_{\rm EI}$ obtained above, RIMS overall detection 
efficiency $E_{\rm RIMS}$ has been estimated to be 
$E_{\rm RIMS}=R\times E_{\rm EI}\sim 7\times10^{-8}$.

The improvement ratio of RIMS detection efficiency with the 808nm laser 
over without it is estimated to be about a factor of $80$ by comparing 
the net ion current of (I) to (III) with the effect of EI ionization 
(II) subtracted. 
This means that the resonance ionization occurs only with the 405nm 
laser, but the ionization is highly effective with the 808nm laser.

The resonance ionization probability of potassium atoms can be estimated 
by solving the rate equations.\cite{RateEq} 
We calculated it for the $\phi2$mm atom beam, estimated from 
the crucible and the orifice geometry, passing through the spherical laser 
spots on the assumption of each laser power and diameter described in 
Section~\ref{exp} and the doppler width of the $T\sim1170$ K 
potassium atoms.
The result was $7.1\times10^{-6}$ with both lasers ON and
$1.3\times10^{-7}$ with only the 405nm laser ON.
The expected improvement ratio of about 60 is a little underestimated
probably because of the transport efficiency through the QMS tube.
The transport efficiency is supposed to decrease with distance from 
the axis of the QMS tube, so the actual improvement ratio is thought to 
be higher due to the larger spot size of the 405nm laser than that of 
the 808nm laser. With these ambiguities, the estimation well reproduces 
the measured ratio.

\subsection{Isotope discrimination}
Isotope discrimination study of potassium has been performed by both 
EI ionization (Section~\ref{EI}) and RIMS (Section~\ref{HFsignal} -- 
\ref{K-40}). 
Here, the channeltron voltage was same as in Fig.~\ref{RIMS-DE}-(a), 
but the crucible heating current was kept a little lower to 6.00 A, 
corresponding to the crucible inner temperature of $T\sim970$ K. 
The mass resolution of the QMS was set higher at the expense of less 
sensitivity to reduce the influence of the tail of the adjacent mass peak.
During the measurements described in Section~\ref{HFsignal} -- \ref{K-40}, 
the filament for EI ionization was OFF. 
The natural isotope ratio of potassium is known to be 
${^{39}{\rm K}}:{^{40}{\rm K}}:{^{41}{\rm K}}=93.2581\%:0.0117\%:6.7302\%$.\cite{NIST}

\subsubsection{Mass spectrum of potassium by EI ionization}
\label{EI}
\begin{figure}[H]
\begin{center}
\includegraphics[height=7cm]{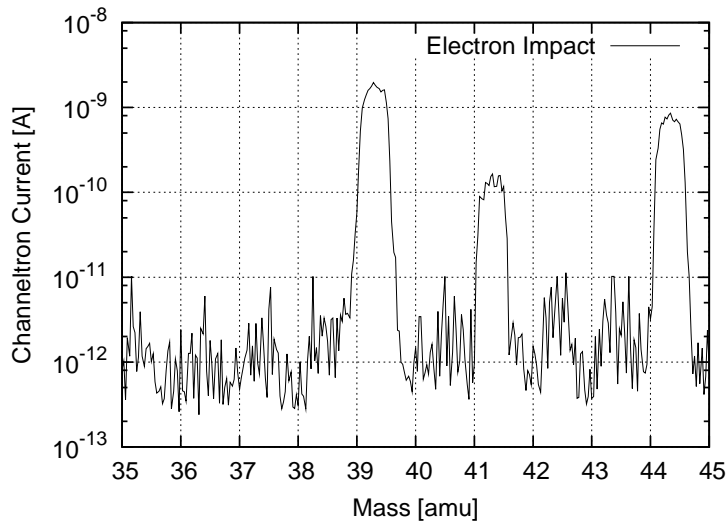}
\end{center}
\caption{Mass spectrum of electron-impact ionization.}
\label{EI-isotope}
\end{figure}

In the electron-impact ionization, isotope discrimination is 
performed only by the mass analysis with the QMS.
Fig.~\ref{EI-isotope} shows an observed mass spectrum around the potassium 
mass region. Though the $^{40}\rm K$ mass peak could hardly be found 
because of the small abundance, the isotope ratio of 
$^{39}\rm K:{^{41}\rm K}=92.3\%:7.7\%$ is obtained from the each 
peak current. 
The deviation from the natural isotope ratio might be 
due to the fluctuation of the channeltron current.

The observed peak at $m/z=44$ is the contribution of ${\rm CO_2}$ 
produced by the decomposition of ${\rm K_2CO_3}$ as shown in 
Eq.~(\ref{decomp1}). 
No atoms nor molecules except for potassium are ionized by the
resonance ionization, so this peak is not observed in RIMS
(See Fig.~\ref{piezo} or Fig.~\ref{40K-peak}).

\subsubsection{Hyperfine splitting of potassium}
\label{HFsignal}
\begin{figure}[H]
\begin{center}
\includegraphics[height=7cm]{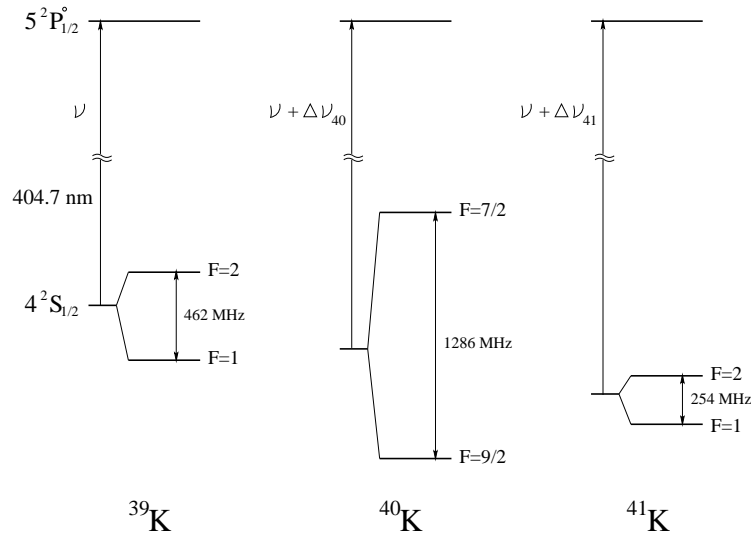}
\end{center}
\caption{Hyperfine structure of the $4\,{^2{\rm S}_{1/2}}$ and $5\,{^2{\rm P}^\circ_{1/2}}$ states of $^{39}\rm K$, $^{40}\rm K$ and $^{41}\rm K$.\cite{Arimondo} Relatively small hyperfine splitting of the $5{\rm P}_{1/2}$ excited state is omitted here. $\Delta\nu_{40}$ and $\Delta\nu_{41}$ are the isotope shifts of the 404.7nm line of $^{40}\rm K$ and $^{41}\rm K$, respectively, with respect to $^{39}\rm K$.}
\label{HFstructure}
\end{figure}

Hyperfine splitting is a splitting of each electron state 
into multiple energy levels caused by the interaction of the 
electron spin induced magnetic dipole moment with the magnetic moment
of the atomic nucleus.
In $^{39}\rm K$ atoms, hyperfine splitting of the $4\,{^2{\rm S}_{1/2}}$ 
ground state makes two energy levels $F=1$ ($-288.6$ MHz) and 
$F=2$ (173.1 MHz) as shown in Fig.~\ref{HFstructure}, resulting in two 
observed resonance ionization peaks.
Here, $F$ is the total angular momentum quantum number including 
nuclear spin.
Hyperfine splitting of the $5{\rm P}_{1/2}$ excited state is relatively 
ignorable.\cite{Arimondo}

\begin{figure}[H]
\begin{center}
\begin{tabular}[b]{c}
\includegraphics[height=7cm]{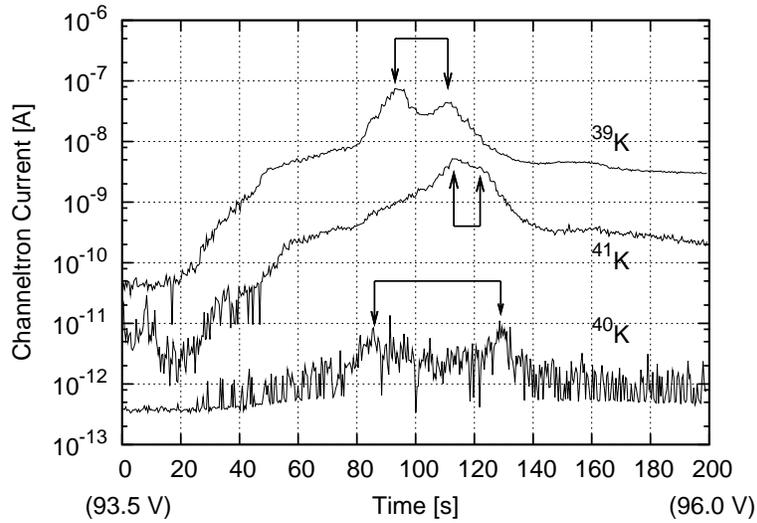}
\end{tabular}
\end{center}
\caption{The observed resonance ionization signal of $^{39}\rm K$, $^{40}\rm K$ and $^{41}\rm K$ by scanning the piezo voltage $V_{\rm piezo}$ from 93.5 V to 96.0 V. The two-headed arrows show the hyperfine splitting of each potassium isotope.}
\label{signal-all}
\end{figure}

Fig.~\ref{signal-all} shows an experimental result of the resonance 
ionization signal of $^{39}\rm K$, $^{40}\rm K$ and $^{41}\rm K$. 
In this figure, the higher peak of $^{39}\rm K$ at the lower $V_{\rm piezo}$, 
or the lower laser frequency, corresponds to the $F=2$ level of the 
$4\,{^2{\rm S}_{1/2}}$ ground state 
of $^{39}\rm K$, and the lower peak to $F=1$ (See also Fig.~\ref{signal}).
The peak current ratio of these two peaks reflects a statistical weight 
of $2F+1$, which is $5:3$. The same is considered to be true for $^{41}\rm K$ 
except for the detailed energy levels of the $F=1$ and $F=2$ level of the 
4s ground state. The two peaks of $^{41}\rm K$ are less clearly observed 
than those of $^{39}\rm K$, because the potassium doppler width of 
about 100MHz is not negligible compared to the narrower hyperfine 
splitting width of $^{41}\rm K$. 
The two peaks of $^{40}\rm K$ correspond to the $F=7/2$ level at the 
lower $V_{\rm piezo}$ and $F=9/2$ level at the higher $V_{\rm piezo}$. 
The peak current ratio of the two peaks of $^{40}\rm K$ is expected to be 
$4:5$.

\subsubsection{Isotope shift of $^{41}\rm K$ with respect to $^{39}\rm K$}
\label{shift}
\begin{figure}[H]
\begin{center}
\includegraphics[height=7cm]{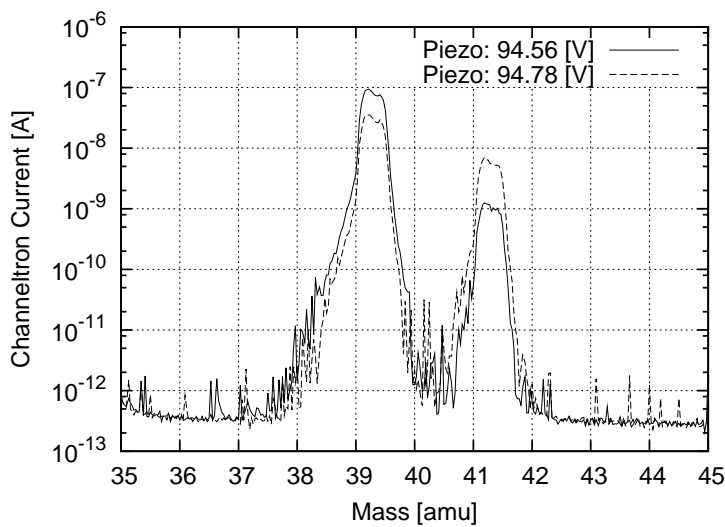}
\end{center}
\caption{The observed isotope shift between $^{39}\rm K$ and $^{41}\rm K$ in terms of the piezo voltage $V_{\rm piezo}$. No peak is observed at $m/z=44$.}
\label{piezo}
\end{figure}

The isotope shift is a slight difference of the transition frequency
(wavelength) between the specific pair of the atomic states of 
two isotopes of the same element. 
The isotope shift of the 404.7 nm resonance excitation line of 
$^{41}\rm K$ with respect to this line of $^{39}\rm K$ can be estimated 
by the observed difference of the piezo voltage $V_{\rm piezo}$ 
at the peak of the measured channeltron current.
Scanning $V_{\rm piezo}$ by hand from 93.5 V to 96.0 V, we found 
the $^{39}\rm K^+$ peak current at 94.56 V and $^{41}\rm K^+$ at 94.78 V.
Both peaks correspond to the resonance excitation line from the 
$F=2$ level of the 4s ground state of each isotope.
Fig.~\ref{piezo} shows a mass spectrum under each piezo voltage 
$V_{\rm piezo}$. 

The observed difference of the piezo voltage was $0.22\pm0.01$ V,
corresponding to the frequency of $(4.4\pm0.2)\times10^2$ MHz 
or the wavelength of $(2.4\pm0.1)\times10^{-4}$ nm.
The error is dominated by the reading error of $V_{\rm piezo}$ 
at the resonant peak
and might be improved with a computer control system of it.
Taking into account the difference of the splitting width of $^{39}\rm K$ 
and $^{41} \rm K$ 4s ground state as shown in Fig.~\ref{HFstructure}, 
the isotope shift is calculated to be $(3.6\pm0.1)\times10^2$ MHz.
Our result is lower than the precise measurement of this isotope
shift using saturation spectroscopy by L. J. S. Halloran {\it et al}.,
\cite{Behr} probably because of the unknown accuracy of the linearity 
between $V_{\rm piezo}$ and the laser frequency, in other words, the 
accuracy of 2 GHz/V as described in Section~\ref{scheme}.

This linearity can be estimated by the known hyperfine splitting width 
of $^{39}\rm K$ $4{\rm S}_{1/2}$ ground state as seen in 
Fig.~\ref{signal-all}. 
The known splitting width of 462 MHz corresponds to the piezo voltage of 
$0.19\pm0.01$ V, so the linearity is obtained to be $2.4\pm0.1$ GHz/V.
With this value, the isotope shift is recalculated to be
$(4.5\pm0.3)\times10^2$ MHz, which is consistent with the data by 
L. J. S. Halloran {\it et al}.\cite{Behr}

As mentioned above, isotopic selectivity can be achieved with 
a CW laser diode in addition to the mass discrimination by QMS.
According to Pulhani {\it et al.},\cite{selectivity} 
the optical isotopic selectivity $\alpha$ for the specific isotope A to
the interfering one B is defined as
\begin{eqnarray}
\alpha = \frac{I_{\rm A}(\nu_{\rm A})/I_{\rm B}(\nu_{\rm A})}{X_{\rm A}/X_{\rm B}},
\end{eqnarray}
where $I_{\rm A}(\nu_{\rm A})$ is the intensity of isotope A at its 
transition frequency $\nu_{\rm A}$, $X_{\rm A}$ is its abundance 
in the sample, $I_{\rm B}(\nu_{\rm A})$ is the intensity of isotope B 
at the transition frequency of isotope A, and $X_{\rm B}$ is its abundance.
Here, the intensity $I$ of each isotope can be replaced by the measured peak 
channeltron current, so the observed $^{39}\rm K$ optical isotopic
selectivity to $^{41}\rm K$ is a factor of 5.
$^{41}\rm K$ optical isotopic selectivity to $^{39}\rm K$ could hardly
be estimated because $^{39}\rm K$ resonance ionization line from the 
$F=1$ level of the ${\rm 4s}_{1/2}$ ground state is close to this 
$^{41}\rm K$ line (See Fig.~\ref{signal}).
It is necessary to reduce the effect of the doppler width for the 
improvement of the optical isotopic selectivity.

The isotope ratio of $^{39}\rm K:{^{41}\rm K}=93.1\%:6.9\%$ is estimated
from the peak current of $^{39}\rm K$ at $V_{\rm piezo}=94.56$ V and 
${^{41}\rm K}$ at $V_{\rm piezo}=94.78$ V in Fig.~\ref{piezo}, which is close to the 
natural isotope ratio and the result of EI ionization described in \ref{EI}.

\subsubsection{Mass spectral peak of $^{40}\rm K$}
\label{K-40}
\begin{figure}[H]
\begin{center}
\begin{tabular}[b]{c}
\includegraphics[height=7cm]{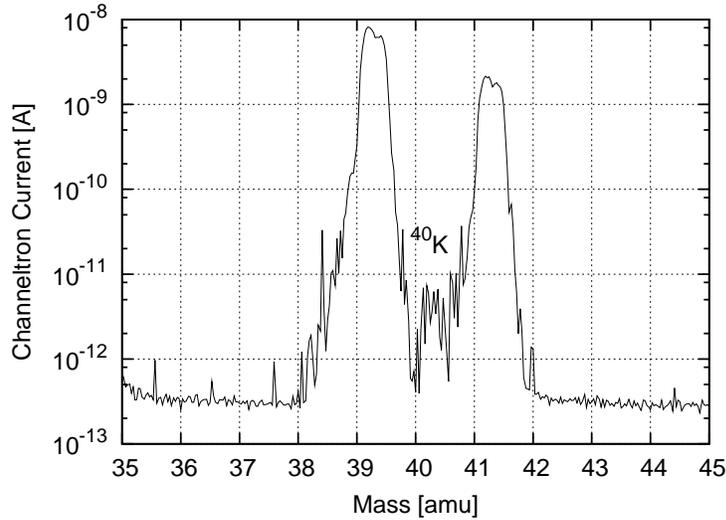}
\end{tabular}
\end{center}
\caption{Mass spectral peak of $^{40}\rm K$ with RIMS.}
\label{40K-peak}
\end{figure}

$^{40}\rm K$ is the naturally-occurring radioactive potassium isotope
which has an abundance of 0.0117\%.\cite{NIST}
In this work, the mass peak of $^{40}\rm K$ has also been observed 
as shown in Fig.~\ref{40K-peak} with a setting of better mass resolution 
(leading to a little lower sensitivity) for the QMS as well as fine-tuned 
piezo voltage to about 95 V. 
This peak is thought to be the resonace excitation signal from 
the $F=9/2$ lower level of the 4s ground state because of the 
higher piezo voltage, in other words, higher frequency of the observed
line than $^{39}\rm K$ or $^{41}\rm K$ (See Fig.~\ref{HFstructure}).

Comparing the peak current at $m/z=39$ and 40, $^{40}\rm K$ optical 
isotopic selectivity to $^{39}\rm K$ is estimated to be about 7.
Taking into account the fact that the isotopic selectivity increases 
with additional resonance excitation processes, higher selectivity could
be achieved by resonance ionization through multiple excited states.

\section{Future Prospect}

Our resonance ionization mass spectrometer is quite effective
for the trace element analysis because of its low noise level of the 
channeltron detector as well as insensitivity to isobaric interferences.
Especially, it can be applicable to the contamination assessment on the 
surface of the semiconductor wafer. 

ICP-MS is mainly used in this field conventionally, however 
it suffers from argon plasma interferences with mass region around 40, 
resulting in relatively worse detection limits of potassium ($^{39}\rm K$)
and calcium ($^{40}\rm Ca$) than other elements.

The detection efficiency of the present system is limited by the 
power of 808nm ionizing laser. Therefore, we are planning to 
increase the 808nm laser power by a factor of 20 using an Apollo 
Instruments model FL-60. 
We expect a detection limit of potassium (K) of comparable to or 
even better than that of ICP-MS with our improved system.

The laser diode is a good device to use for the resonance excitation 
because of its narrow linewidth and relatively low price. However, 
it is not applicable to the analysis of other chemical elements for the 
limited variable wavelength range.
For the contamination assessment of a variety of impurities on the wafers, 
a variable wavelength dye laser can be used at a higher cost.

\section{Conclusion}
Trace element analysis of potassium gas atoms has been performed by 
our own resonance ionization mass spectrometer consisting of two 
laser diodes for resonance ionization and QMS for mass analysis.
The overall detection efficiency is currently estimated to be 
$7\times10^{-8}$. 

Isotope discrimination test has also been performed with a high 
resolution QMS setting to verify the isotope shift and the 
optical isotopic selectivity. 
The isotope shift between $^{39}\rm K$ and $^{41}\rm K$ in the 
404.7nm resonance excitation line is estimated to be  
$(4.5\pm0.3)\times10^2$ MHz compared to the known hyperfine splitting 
width of $^{39}\rm K$ $4{\rm S}_{1/2}$ ground state.
The obtained optical isotopic selectivities of $^{39}\rm K$ to 
$^{41}\rm K$ and $^{40}\rm K$ to $^{39}\rm K$ are about 5 and 7, 
respectively. 
A clear hyperfine splitting of the $4{\rm s}_{1/2}$ ground level 
was observed in each potassium isotope. 

We are planning to apply RIMS to the contamination assessment on the 
surface of the semiconductor wafer. 
The detection limit of potassium impurities can be improved to be 
comparable to or even better than that of a conventional method of 
ICP-MS by increasing the 808nm laser power by a factor of 20.
Also, a wavelength-tunable dye laser to cover the wide range of 
resonance excitation wavelength enables us to perform contamination
assessment of a wide variety of impurities in addition to potassium.

\section*{Acknowledgment}
This research is supported by Grant-in-Aid for Exploratory Research, 
the Japan Society for Promotion of Science, by the Research Center for the 
Early Universe, School of Science, University of Tokyo, and also by the 
Global COE Program "the Physical Sciences Frontier", MEXT, Japan.

\end{document}